\documentclass[%
mathleft,%
]{an}
\usepackage{graphicx}
\usepackage{times}
\begin{document}

\Pagespan{1}{}
\Yearpublication{2011}%
\Yearsubmission{2011}%
\Month{1}%
\Volume{999}%
\Issue{92}%

\title{Accretion in active galactic nuclei \\
     and disk-jet coupling }

\author{Bo\. zena Czerny\inst{1,2}\fnmsep\thanks{Corresponding author:
  \email{bcz@cft.edu.pl}}
\and  Bei You\inst{2}
}
\titlerunning{Accretion in AGN and disk/jet coupling}
\authorrunning{B. Czerny \& B. You}
\institute{
Center for Theoretical Physics, Al. Lotnikow 32/46, 02-668 Warsaw, Poland
\and 
Copernicus Astronomical Center, Bartycka 18, 00-716 Warsaw, Poland}

\received{XXXX}
\accepted{XXXX}
\publonline{XXXX}


\abstract{%
  We review the current state of understanding how accretion onto a black 
  hole proceeds and what the key elements needed to form relativistic jets are.
Theoretical progress is severely undermined by the lack of
thorough understanding of the microphysics involved in accretion discs and
in the formation of jets, particularly in the presence of strong magnetic
fields. Therefore, all proposed solutions are still models that need to be
validated by observational constraints.}

\maketitle

\section{Introduction}
The formation of relativistic well collimated jets extending well beyond the host galaxy 
is a fascinating property of active galaxies. The subject is not new: the first paper listed in ADS which contains the term 'jet'
in the title is the paper by Baade (1956), discussing the polarization of the jet in M87. Jet formation and jet stability was already studied not much later (Bisnovatyi-Kogan, Komberg \& Fridman 1969). Since those early days, observational progress has been enormous thanks to the availability of new instrumentation covering broad band spectra and capturing images in various energy bands from radio through optical to high energy gamma rays, and the multiwavelength campaigns shed light onto the complex variability pattern. However, understanding of the details of the physical mechanisms behind the jet formation progresses more slowly.

\section{Broader view - objects with jets}

We know from observations that jets are very common. Jets form around protostars, galactic X-ray binaries, gamma-ray bursts and active galactic nuclei (AGN). All these systems form both uncollimated winds and collimated jets. A jet was even discovered in the case of the Tidal Disruption Event of a star by a black hole (Bloom et al. 2011;  Cenko et al. 2012). Notable exceptions are cataclysmic variables - these systems form winds  but do not form jets. These systems may thus be used for testing jet models.

Jets in these systems show a range of properties. Protostar jets are slow while the AGN jets are relativistic. Since the subject of this review are AGN jets, we will be interested only in relativistic jets. Such jets clearly form when the central object is a black hole (the case of AGN and gamma-ray bursts). It is still under discussion whether neutron stars are also able to form relativistic jets. 
For some time Circinus X-1 was considered a nice example of a relativistic jet in a neutron star system (Sell et al. 2010; Miller-Jones et al. 2012). However, a recent study suggests that the observed emission is better explained as being due to a supernova remnant instead of a relativistic jet (Heinz et al. 2013). If so, it seems that the existence of a black hole is required to form a highly relativistic outflow. 

On the other hand, it is not clear whether all black holes always produce some form of relativistic jet. The studies of the Galactic black holes clearly show that strong jets accompany certain spectral states. Galactic black holes evolve on short timescales, and X-ray novae in a matter of years can cover all luminosity states from quiescence through outburst and back to quiescence. Jet formation clearly accompanies the hard states, and the jet line on the X-ray hardness ratio vs. luminosity diagram marks the border between the soft and hard states (Fender et al. 2004). In addition, a jet in the lower part of the diagram (lower states) is a kind of stationary jet, while jets in the upper part of the diagram form blobby outflows. When the source enters the soft disk-dominated state, the presence of the jet is not seen, although there is always a hard X-ray tail to the spectrum and the emission may come either from some form of corona or from a weak jet (e.g.  Del Santo et al. 2013). 

AGN are also divided into radio-loud and radio-quiet. Spectacularly well-collimated jets in some radio galaxies are observed in radio, optical and X-ray band. However, radio-quiet sources are not radio-silent (see e.g. the composite spectra of  Laor et al. 1997). Radio-quiet sources are generally unresolved on the radio maps, and, if marginally resolved (see e.g. Giroletti \& Panessa 2009 and Khard et al. 2015 for VLBI observations of a few Seyfert 1 galaxies) the size is so small that the nature of emission is difficult to establish, so the presence of a jet is possible but not proved. 

\section{Accretion disk models}

\subsection{Keplerian disk}

The best studied disk model was proposed by Shakura \& Sunuaev (1973). The full version of the model is based on the alpha-viscosity parameter, i.e. the scaling of the stress with the total pressure. However, if we consider only a stationary solution, and the disk is optically thick, the solution does not depend on the viscosity, and it is parametrized by the black hole mass, accretion rate and the inner radius where stress vanishes. The Newtonian description of the disk around a black hole is done by setting the inner radius at $3 R_{Schw}$, or $6 R_{g}$. 
\begin{equation}
F(r) = {G M \dot M \over r^3}(1 - \sqrt{6 R_g \over r}),
\end{equation}
where $F(r)$ is the radiation flux at radius $r$, $M$ is the black hole mass, $\dot M$ is the accretion 
rate.

The description is so simple because in this model the specific angular momentum of a fluid element determines the radial position in the disk and the angular momentum transfer outward and the disk emissivity is specified by the designated accretion rate. The model was soon generalized to full General Relativity (Novikov \& Thorne 1973), and then the inner radius is provided by the spin-dependent location of the Innermost Stable Circular Orbit (ISCO). The stationary model is again uniquely given by the black hole mass, accretion rate and a black hole spin. The position of  ISCO changes from $1 R_{g}$ for a maximally rotating black hole to $12 R_{g}$ for a  maximally counter rotating one. Closer to the horizon the matter flows in supersonically with negligible dissipation (Muchotrzeb \& Paczynski 1982). The model simplicity is due to the fact that the disk material moves approximately on the Keplerian orbits, the disk is geometrically thin,  and radial pressure gradients are negligible so the solution is fully determined by conservation laws of mass and angular momentum. The disk emission is also uniquely determined if the emission can be locally treated as a black body emission. 

Any departure from the assumption of a stationary, Keplerian, optically thick disk leads to enormous complications due to the explicit dependence of the emitted radiation flux on the disk structure which in turn depends on the spatial distribution of dissipation through the action of viscosity.  Even simple color corrections to the black body spectra due to electron scattering are highly uncertain (e.g. Davis \& Hubeny 2006, Done et al. 2012, You et al. 2015a). The big issue is the time evolution and the disk stability. Simple estimates of the thermal and viscous timescales in the Shakura-Sunyaev model show this clearly through the explicit dependence on the viscosity parameter $\alpha$:
\begin{equation}
t_{th} = {1 \over \alpha} t_{dyn}; ~~~t_{visc} = ({r \over H})^2 t_{th},
\end{equation} 
where the dynamical (Keplerian) timescale $t_{dyn} = {r^3/G M}$. Actually, soon after they were developed, the Shakura-Sunyaev disks were shown to be thermally and viscously unstable. For some years the impression that unstable disks cannot exist strongly affected the model developments. Now it seems that the instabilities actually serve as explanations of some of the variability patterns. The partial ionization instability is responsible for the outbursts  of dwarf novae and X-ray novae, and the radiation pressure instability is responsible for the heartbeat states in GRS 1915+105 (e.g. Janiuk et al. 2002), and IGR J17091–3624 (Janiuk et al. 2015) and (likely) for GPS sources reactivation (Czerny et al. 2009).  Observations support the intermittent character of significant fraction of the small scale core-dominated sources (e.g. Marecki et al. 2006, Kunert-Bajraszewska et al. 2006, Shulevski et al. 2015). In general, the are direct arguments for important effects of the time evolution in the observed radio-sources (see Kunert-Bajraszewska et al. 2010) so in modelling we have to go beyond the stationary models.

A major step forward in the disk models came with the understanding of the physical nature behind the viscosity: Balbus \& Hawley (1991) rediscovered the known (but not in this context) magnetorotational instability (Velikhov 1959, Chandrasekhar 1961) and explained that the small-scale magnetic field provides an effective mechanism of angular momentum transfer. Current numerical simulations indicate that this
mechanism roughly corresponds to the value of the viscosity parameter $\alpha$ in the range 0.03 - 0.3 (Penna et al. 2013). Observational constraints are roughly consistent with this range in the case of AGN (Siemiginowska \& Czerny 1989; Starling et al. 2004; Xie et al. 2009) although values lower by more than an order of magnitude are found by Kelly et al. (2009).

The understanding of the viscosity mechanism does not mean that all the problems are now solved in a satisfactory way when we use MHD simulations. It is best illustrated by the history of numerical studies of disk stability. Early attempts showed that the radiation pressure instability is not present in the disk (Turner 2004, Hirose et al. 2009ab), and this was explained either as a stochastic damping of the instability (Janiuk \& Misra 2012) or as a delay between the field generation and dissipation (Ciesielski et al. 2012). However, the latest simulations show (Jiang et al. 2014) that the radiation pressure instability does exist! This result required a new code (Athena), and more importantly, a much larger radial grid. Also, the partial ionization instability was confirmed with the lastest simulations by Hirose et al. (2014).

The Keplerian disk model makes simple predictions for the overall spectrum. The multicolor disk is expected to form a single peak, with roughly exponential decay at high energies (far UV/EUV) related to the maximum value of the disk temperature, and a $ F_{\nu} \propto \nu^{1/3}$ asymptotic behaviour at longer wavelengths (optical/IR). Simple stationary Keplerian disks serve many purposes like broad band fitting of the quasar optical/UV spectra (see e.g. Capellupo et al. 2015), or explaining the formation of the Broad Line Region (e.g. Czerny \& Hryniewicz 2011) well.  They do not, however, produce jets.


It is now widely believed that the relatively well studied Keplerian disk models are not adequate
to form relativistic jets. High ram pressure is needed to accelerate the jet near its base, in the vicinity of the black hole horizon. This is why the highest Doppler factors are observed in the case of gamma ray bursts. Keplerian disks are based on the assumption of the radial pressure being negligible so Keplerian disks themselves are not suitable for launching relativistic jets.

\subsection{Non-Keplerian disks}

Non-Keplerian disks come in a large variety since these models are strongly dependent on the underlying assumptions concerning the angular momentum distribution, optical depth, emissivity, ouflow pattern and large scale magnetic fields. 

One family of disk models appeared as a small deviation from a spherically-symmetric Bondi (1952) flow; a small but non-negligible amount of angular momentum leads to the formation of additional critical/transonic points (Abramowicz \& Zurek 1981). In spherical accretion the Bondi flow becomes supersonic quite far from the black hole (although the Bondi radius depends strongly on the polytropic index of the gas) while in low
angular momentum flows the sonic radius depends on the angular momentum and may be located inside the ISCO. Such models do not need viscosity because the inflow proceeds even without the need for angular momentum transfer. These models are still under examination in the context of inactive galactic nuclei where the angular momentum of accreting matter is not necessarily high (e.g. Moscibrodzka et al. 2006; Das et al. 2015). Polish oughnuts form the other branch of the same family of models: they are  non-viscous tori around black holes (e.g. Qian et al. 2009).

The second family of disk models are sub-Keplerian disks, where the local angular momentum is a fraction of a Keplerian angular momentum. These models require viscosity for the matter to accrete since the incoming material faces the angular momentum barier. However, they are geometrically thick, with strong radial pressure gradients and accretion proceeding much faster than in the Keplerian disk, roughly on the thermal timescale.  

Such disks can be optically thick, and these disks are known as slim disk models (Abramowicz et al. 1988; for a review, see Lasota 2015). They provide an appropriate descriprion for high angular momentum accretion flow, where part of the energy is advected inwards due to a timescale for accretion much shorter in comparison with the timescale for photon propagation from the disk interior towards the disk surface. A hint that this could happen was contained already in the paper by Meszaros (1975)  describing the spherical accretion onto a black hole including high accretion rates.

Such disks can be also optically thin, and this happens when the accretion rate is low. The first such model
was proposed by Ichimaru (1977), and broadly popularized under the name of ADAFs (Advection-Dominated Accretion Flow) by Narayan \& Yi (1994). Further developments include ADIOS (Advection-Dominated-Inflow-Outflow; Blandford \& Begelman 1999), CDAF (Convection-Dominated Advection Flow; Quataert \& Gruzinov 2000), MAD (Magnetically-Arrested Disks; Narayan et al. 2003). Sometimes they are generally referred to as RIAF (Radiatively-Inefficient Accretion Flow), although that can include the spherical Bondi flow and the low angular momentum flows discussed above.  The radiation emitted by RIAF is generally assumed to be a second order effect  so the time-dependent modelling can concentrate on the dynamics. On the other hand, the
actual level of emissivity is very difficult to estimate reliably. Starting from the early models it was recognized that close to the black hole a two-temperature plasma developes, with ions close to the virial temperature (consistent with large geometrical disk thickness) and electron temperature saturated roughly at 100 keV. The electron temperature may certainly be limited by the possibility of efficient pair-creation as first recognized by Bisnovatyi-Kogan, Zeldovich \& Sunyaev (1971) but efficient cooling may also happen through the small fraction of the non-thermal electrons emitting synchrotron radiation. Recent results from NuSTAR supports this view but they are not yet accurate enough to firmly distinguish between the two saturation mechanisms (Fabian et al. 2015). In addition, the flow of energy from ions to electrons is not well established; Coulomb coupling provides the minimum coupling but more complex plasma processes may enhance coupling, and part of the energy may go directly to electrons.

The spectral predictions for various RIAF models depend on specific assumptions but generally the emission is expected to consist of two separate components. One component, visible in the radio/IR/optical part is due to synchrotron emission of the electrons and the second component visible in X-rays is due to
Comptonization by the same electrons. The overall prediction, thus, resembles the jet emission but emission from 
the disk is not relativistically boosted. 

\subsection{Coexistence of different flow patterns} 

We see directly from the AGN spectra that the accretion in most AGN cannot be described with a single model listed above. Spectra of radio-quiet AGN are dominated by the Big Blue Bump extending from the optical/UV to soft X-ray band. The optical part of the spectrum is well described by a Keplerian disk (see Capellupo et al. 2015 for very nice fits to several individual AGN to the XSHOOTER data).  However, in the X-ray band we see additional emission both in the soft X-ray band and in the hard X-ray band. In the case of
lower luminosity sources, like radio galaxies with double emission line profiles the cold Keplerian disk seems to be present at large radii only, and the innermost flow is of the RIAF type. Very low luminosity sources, like
the extremely faint Sgr A* - the Milky Way center - do not show any presence of a cold disk. 

There are thus many ad hoc models illustrating various possibilities of the relative setup of the cold disk and a hot plasma. The same models apply for AGN and the black holes in binary systems.  Sandwich type geometry, with the hot optically thin plasma surrounding cold accretion disk has been discussed in a number of papers (Bisnovatyi-Kogan \& Blinnikov 1976; Galeev, Rosner \& Vayana 1979; Begelman, McKee \& Shields 1983; Haardt \& Maraschi 1991), and this corona can be either anchored to the disk by the magnetic field, like the solar corona, and heated by magnetic field reconnection, acoustic waves and irradiation, or the corona itself can accrete (\. Zycki et al. 1995; Chakrabarti \& Titarchuk 1995) and dissipate the gravitational energy. In another approach the transition between the cold disk and the corona can be predominantly radial, with the hot flow usually in the inner part (Rees 1982; McClintock et al. 1995; for recent papers se e.g. Qiao et al. 2013) but the flow can also start as a hot flow at large radii (e.g. Liu et al. 2015). 

The physical description of the transition between the corona and the disk is still far from being satisfactory. The mass exchange between the disk and the corona (i.e. evaporation and condensation) was introduced by Mayer \& Mayer-Hoffmeister (1994) in the context of the cataclysmic variables, and was later adopted in the description of accretion onto black holes (e.g. Rozanska \& Czerny 2000; Mayer \& Pringle 2007). The key ingredient in this process is electron conduction, on top of the radiative heating/cooling. The process efficiency depends on the presence of a magnetic field (Meyer \& Meyer-Hofmeister 2002). This line of studies implies that there may be a secondary formation of a cold, mostly passive thin disk close to the black hole, separated radially from the outer cold disk (Meyer-Hofmeister \& Meyer 2011). In addition, not only electron conduction but also ion conduction should be included but this issue has not been studied yet apart from preliminary attempts (Dullemond \& Spruit 2005). Time-dependent MHD simulations never produce such a feature as a secondary passive inner disk but they likely lack the appropriate spatial resolution and/or proper cooling description. Therefore, we are still far from the possibility to calculate the exact flow pattern from the assumed outer reservoir of the plasma and when interpreting the data we generally have to resort to some parametric treatment. 

The coexistence of the hot plasma and a cold disk is an attractive scenario for accretion, both from the point of view of modelling broad band spectra and from the point of view of jet formation.

\subsection{Fundamental Plane}

Which of the numerous models correspond to the observational data for a given object, and which are suitable for jet production? In principle, it is possible to determine that directly from observations. Merloni et al. (2003) discovered the tight relation between the radio luminosity of radio-loud AGN and the  combination of the X-ray luminosity and the black hole mass in the logarithmic plane
\begin{equation}
 \log L_R = 0.6 \log L_X + 0.78 \log M + 7.33,
\end{equation}
referred to as the Fundamental Plane. The coefficients in this relation can be compared with theoretical 
predictions based on a disk model and the assumption of a jet 
The relation extends down to small AGN masses (Gultekin et al. 2014) but disk-dominated radiatively efficient flows follow a different track (Fender \& Gallo 2014; Dong et al. 2014). This is not surprising since the two disk models are different. However, Merloni et al. (2004) warn against trying to use the inversion of the approach and determine the disk properties from the slope of the Fundamental Plane as the scatter is still quite large, and in addition the variability in the disk and filtered variability of the jet contribute to the scatter (Lin et al. 2015). 

\section{Jet production: large scale magnetic fields}

A geometrically thick accretion flow, either in the form of a RIAF, or coronal flow on the top of the cold thin Keplerian disk, or perhaps a highly super-Eddington slim disk seems necessary to produce a relativistic jet but they are not automatically sufficient conditions. MRI operating in these disks provides the viscosity but does not lead to the formation of a large scale magnetic field which is also a necessary ingredient. Such large scale magnetic fields are proposed to be generated through two possible scenarios: (i) magnetic field line dragging (ii)
generation in situ by a cosmic battery. 

The first idea dates back to Bisnovatyi-Kogan \& Ruzmaikin (1974), and was further elaborated e.g. by Narayan, Igumenshchev \& Abramowicz (2003), and Sikora \& Begelman (2013). As Sikora \& Begelman (2013) point out, only a hot, geometrically thick (in fact, an almost spherical) flow is able to provide the efficient dragging. Simulations support the phenomenon of dragging but the problem is that the effect strongly depends on the initial configuration of the magnetic field which in turn modifies the flow pattern (Penna et al. 2010).
Depending on the setup, we can have relativistic jets (e.g. McKinney et al. 2014) or non-relativistic jet with additional efficient wind outflow (e.g. Yuan et al. 2015). Similar results come all the groups: solutions with strong un-collimated outflow and solutions with a jet-like structure are found (Ohsuga et al. 2005; Takeuchi et al. 2010; Ohsuga \& Mineshige 2011; Takahashi \& Ohsuga 2015). 

The field dragging leads to the accumulation of the magnetic field close to the black hole horizon and formation of Magnetically Arrested Disks (MAD; Narayan et al. 2003).

The second idea dates back to Contopoulos \& Kazanas (1998) further developed e.g. in Contopoulos et al. (2015). Here the magnetic field is generated due to the difference of the radiation pressure acting on electrons and protons, which then leads to the relative motion of electrons and protons, subsequent electric field generation and finally generation of the magnetic field. The mechanism creates large scale poloidal magnetic loops. It is not clear whether the efficiency of this mechanism is high enough to produce powerfull jets. There is also the concept of the Biermann battery (Biermann 1950; see also Rhanna 1998), which creates a magnetic field due to the acceleration operating in a rotating plasma, and its most recent developments have been discussed by Schoeffler et al. (2014). This line of research is much less advanced than the field dragging but the mechanism might be very attractive since the result is local, does not depend on the outer boundary conditions and thus implies likely similarities between the galactic sources powered by the companion star and AGN powered by the surrounding medium. Field dragging in these two types of sources would instead lead to significant diferences in the jet formation efficiency.

Recently Parfrey et al. (2015) proposed that the large scale magnetic field is not actually necessary to produce a jet due to the role played by magnetic field reconnection. More studies in this field are clearly needed.

\section{Jet production: source of energy}

Strong magnetic fields govern the jet formation and confinement, ram pressure provides the momentum but we have to answer the question about the source of energy. There are two possibilities: the energy can come from the accretion flow or it may come from the rotational energy of the central black hole. If the field dragging  is the operating mechanism for the large scale magnetic field production, and if as a result of accumulation of the magnetic field close to the black hole horizon a MAD disk develops  we have the required ram pressure (as matter comes to a halt close to the horizon) but we may not then have enough energy to actually power the outflow (Marek Sikora, private communication). In this case the extraction of rotational energy from the black hole is necessary. It is generally refered to as the Blandford-Znajek process, and this scenario is relatively well developed.

There are also alternatives. For example, Singh et al. (2015) propose that  fast  magnetic field reconnection takes place in the disk corona and this leads to plasmoid ejection providing a description of jet formation in high Eddington ratio sources. 

\subsection{Blandford-Znajek process}

Jet energy can come from the rotational energy of a black hole. It was proposed by Penrose that this energy can be extracted
through a decay of particles in the ergosphere and escape of one of the decay products to infinity (Penrose 1969). This extraction is also possible through the action of the elecromagnetic fields as proposed by Blandford \& Znajek 1977  (for a review, see Lasota et al. 2014).

High spin above $\sim 0.5$ is required for high efficiency of the process, as indicated by the formula (Garofalo 2009)
\begin{equation}
L_{BZ} = {1 \over 32} B_H^2 r_H^2 a^2 c,  
\end{equation}
where $B_H$ is the magnetic field, $r_H$ is the black hole horizon radius, $a$ describes the black hole spin and $c$ is the light speed. Here we dropped the term connected with the angular momentum of observers measuring zero electric field. The formula gives high efficiency for large $a$, both positive (co-rotating disk) and negative (counter-rotating disk), Garofalo (2009) actually argues that a counter-rotating disk is more efficient by a factor of 2. Numerical GR MHD simulations support the high efficiency of the mechanism (e.g. Tchekhovskoy, Narayan \& McKinney 2010; McKinney et al. 2014) but because of model limitations the final test should come from observations, i.e. high spin sources should have jets, and more importantly, low spin sources cannot have strong jets.  

\section{Spin measurements in AGN}

There are several methods of spin measurement in AGN. Some of them were already used in the case of several sources, while others await appropriate observational opportunities (for a review, see Brenneman 2013). 

The oldest method is based on modelling the relativistically broadened Kalpha iron line (Iwasawa et al. 1996). The method is based on the dependence of the iron line shape on the spin, but the actual measurement is quite complex since it has to account for the line emissivity profile in the disk, the effect of the warm absorber, and reflection from a distant medium, as discussed by Reynolds (2013). The iron line is produced as a result of the irradiation of the inner cold Keplerian disk by a hot plasma, so the location of the X-ray emitter and the ionization state of the disk affect the line emissivity in a complex way. Nevertheless, the method has been applied to a number of AGN as well as galactic black holes with some success (see reviews by Brenneman 2013 and by Reynolds 2014). The spin measured with this method is always larger than 0.5 although the measured AGN are radio quiet. This has implications for galaxy evolution since the angular momentum of accreting material is related to the kinematics of the host galaxy (Sesana et al. 2014). On the other hand, there may be an observational bias since the iron line expected from a fast rotating black hole is stronger and it is thus more easily measured.  

\begin{figure}
\includegraphics[width=\linewidth,height=60mm]{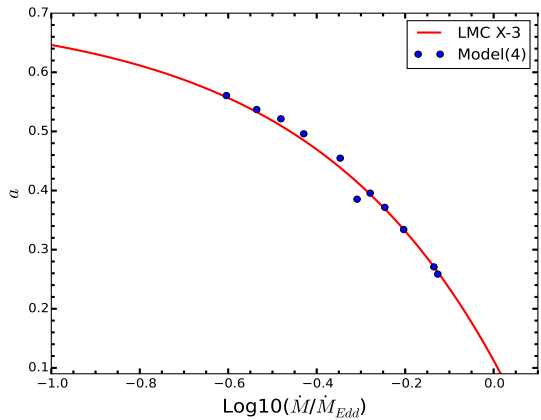}
\caption{The apparent decrease of the spin with luminosity in LMC X-3 (dots) obtained from the disk fitting method can be well modeled as a result of strong outflow (continuous line) from the innermost few gravitational radii (You et al. 2015a).}
\label{bei}
\end{figure}

\begin{figure}
\includegraphics[width=\linewidth,height=60mm]{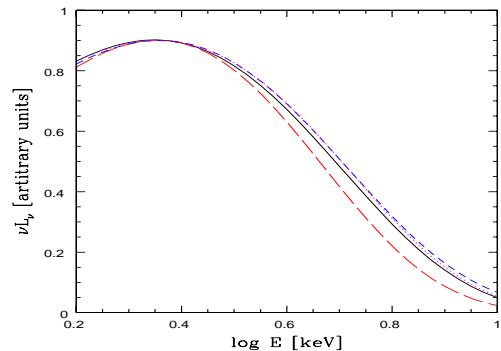}
\caption{The shapes of the continuum disk emission for a disk in a binary system extending to 
ISCO (dotted magenta line), disk with inner cut-off (black continuous
line), disk with cut-off after Comptonization by a spherical wind with optical depth $\tau = 5$, wind temperature 
$T = 1keV$ (red long dashed line) or $\tau = 0.05$, $T = 100keV$ (blue short
dashed line). Model parameters were adjusted so that the position of the maximum in the spectra coincide. Models cannot be distinguished with RXTE data (You et al. 2015a).}
\label{slope}
\end{figure}

The second method is known as the Continuum Fitting method and it is based on modelling the cold disk continuum. The maximum temperature of the accretion disk and its inner radius (set by ISCO) strongly depend on the spin. This method was first applied to galactic sources (see McClintock et al. 2014 for a review) but was recently used also for quasars (Czerny et al. 2011, Capellupo et al. 2015).  Spin values obtained with this method (again, for radio-quiet sources) span the whole range, from counter-rotating disk with $a = -1$ through moderate rotators to high spin values $a = 0.998$. The difficulty with this method lies in a need for independent black hole mass measurement (otherwise the fit is not unique) as well as in the issue of color correction to the disk effective temperature and the possibility of an outflow.  The last two difficulties were  recognized in the case of galactic sources when the fitted value of the spin decreased with the increase of the luminosity in a studied source (McClintock et al. 2006), and improvement in the disk model (slim disk approach) did not remove the problem (Straub et al. 2011). An outflow can well mimick this phenomenon.  In Fig.~\ref{bei}) we show the apparent decrease of the spin which came from a standard disk fitting for LMC X-3 (Straub et al. 2011) in the case when the the color correction to the disk temperature came with the BHSPEC model in XSPEC (points). We tried to reproduce this behaviour with models of a radius-dependent outflow but a parametrization not containing explicit dependence on luminosity was not appropriate. On the other hand, the simplest parametrization of a disk with an luminosity-dependent inner cut-off worked well (Fig.~\ref{bei}, continuous line), and we were able to fit the data well with such a model. In the case of an outflow, the outflow can/should modify the disk spectrum depending on the outflow colllimation and plasma temperature. In principle, high quality data should be able to differentiate between different scenarios, including the specific value of the inner cut-off since the relativistic effects depend both on the black hole spin and cut-off radius in a subtly different way. However, in practice the differences in the models are small when all the parameters are set in such way as the spectra have their maxima in the same position (see Fig.~\ref{slope}), and at high energies there is still the additional hard X-ray power law. 

A modification of this method can apply if the position of the maximum is not in the visible range but the information about the X-ray spectral shape is available. Since the normalization and the slope of the X-ray emission from the corona depends on the disk emission through heating/cooling balance the broad band fits can still return the spin value (You et al. 2015b).  

The following methods might be applicable in the future.

One method is related to the high frequency quasi-periodic oscillations (QPO) observed in a number of black holes in binary system and firmly detected in two AGN - RE J1034+396 (Gierlinski et al. 2008) and MS 2254.9-3712 (Alston et al. 2015). The QPO is a temporary feature and apparently not easy to spot in AGN. The method is basically simple since those oscillations are believed to be related to the orbital frequency at ISCO. However, in binary black holes the QPO frequency has been changing with the source luminosity state which implies that radiative processes can modify the position of the inner radius. Still, perhaps the maximum value in a given source (in the case of galactic sources when we can observe several luminosity states) may provide the appropriate measurement. In AGN luminosity states last thousands of years so this uncertain factor may limit future applicability of this method to AGN. 

Still another method proposed by Schnittman and Krolik (2009) uses the dependence of the polarization degree and polarization angle on energy based on the predictions of a thin disk and the dependence of the ISCO on the spin. 

Imaging the Event Horizon Shadow is in attractive option for weakly active galactic nuclei. This can be done in the near future in the mm band (Doeleman et al. 2008), at least for Sgr  A* and for M87, nuclei with the largest angular size of the black hole horizon (tens of micro arc seconds). The image shape depends on the black hole spin. 

Finally, it is possible to measure directly the angular momentum of incoming photons apart from their polarization. Technical feasibility of the measurement, with the possible applications to black hole spin measurement was discussed by Thide et al. (2007), and in more detail by Tamburini et al. (2011).    

The comparison of the spin measurements based on two different methods done for X-ray binaries showed  that the results are occasionally contradictory (see Steiner et al. 2011 and a review by Brenneman 2013).

The work done for X-ray binaries indicate a rather moderate correlation between the jet and a spin (Miller et al. 2009); objects with highly relativistic jets have higher value of spin measured and Cyg X-1  has very low value of spin although it has the jet velocity above 0.6 of the light speed. So while the spin paradigm for jet power is attractive (Meier 2001) it is still awaiting broad observational justification.

\section{Jets in GPS-CSS sources}

Gigahertz peaked-spectrum (GPS) and compact steep-spectrum (CSS) sources have, in general, jets well confined within their host galaxies, with a few exceptions like 300 kpc jet in PKS 1127–145 (Siemiginowska et al. 2007). Small jet sizes imply that either these sources are young or the jet propagation is frustrated by the dense interstellar medium (see Callingham 2015, Orienti 2016 and the references therein). Despite the numerous observational studies, the issue is not set. Accretion disk instabilities discussed in Sect. 3.1 form an attractive scenario if the activity is intermittent and the amount of accreting gas large enough to allow completion of a number of limit cycles. Direct studies of accretion disks in these sources in general is not easy since they are significantly obscured in the optical/UV band. However, black hole masses were measured for a smaple of sources by Wu (2009) and rather high values of the Eddington ratio in these sources were consistent with disk instability scenario. 

\section{Radio quiet AGN and jets}

The radio loud/radio quiet AGN dichotomy has been strongly disputed during the last years. At first the division was obvious since some AGN show extended well resolved jet feature while others do not, and the radio loud sources are thousands of times brighter in radio than radio quiet ones. However, the broad band quasar spectra do not show large differences between radio loud and radio quiet sources (radio-loud quasars have somewhat redder slopes in UV and somewhat harder slopes in hard X-rays), and the amount of energy emitted in the radio band in these sources is small (even for radio-loud objects) in  comparison with the Big Blue Bump. Lower luminosity AGN, with weak or absent Big Blue Bump are always radio-loud since the radio loudness is traditionally defined as the ratio of the flux at 5 GHz to the flux at  optical B band.

Therefore a natural question appears whether the radio-quiet sources are unable  to produce relativistic jets
or the jets are too weak to propagate significantly. Ghisellini et al. (2004) argued for an aborted jet in radio sources due to too low initial jet velocity. The radio structures in Seyfert galaxies are frequently resolved up to a distance of 1 kpc (Gallimore et al. 2006). However, a fraction of Seyfert galaxies do not show relativistically broadened iron line and thus these sources can have an inner optically thin flow. The issue was recognized by Kadler et al. (2005) but collecting the VLBI data for sources with relativistic iron line is apparently difficult. Radio structures in radio quiet quasars are unresolved. 

Garofalo, Evans \& Sambruna (2010) proposed that radio-loud AGN contain counter-rotating disks and radio-quiet AGN contain co-rotating disk. While for majority of the sources it may work, the giant radio quasars with radio structures larger than 0.7 Mpc (Kuzmicz \& Jamrozy 2012) likely pose a problem. Their huge size of the radio-structure implies long-lived stable accretion which should over time change the counter-rotating scenario to co-rotation (see e.g. Sikora 2009). 

\section{Conclusions}

Formation of relativistic jets remains a challenging problem. Basic ingredients required to power jets are probably well identified but the exact mechanism, and the details, are still unspecified. This is not surprising since the time-dependent processes in which magnetic fields play a strong role are very complex.Numerical MHD codes used to simulate the flow close to the black hole horizon improve, and include the General Relativity effects, radiative cooling and start to capture the magnetic field reconnection. However, they cannot resolve all the microscopic effects covering at the same time the required enormous range of spacial and temporal scales. It is not yet possible to determine the plasma conditions close to the black hole self-consistently knowing just the mass supply.  Fortunately, we
can expect a flow of new data, including the broad band monitoring of selected sources. Combined effort
to improve AGN statistics and successful spin fitting can help to constrain the jet formation mechanisms observationally. On the other hand, more detailed high resolution observations of individual objects give constraints on the magnetic field, jet content and substructure. This, in confrontation with the models, should give a better insight into the jet formation process.

\acknowledgements
  This research was supported by Polish
National Science Center grants No. 2011/03/B/ST9/03281,\\
2013/08/A/ST9/00795, and by Ministry of Science
and Higher Education grant W30/7.PR/2013. It received
funding from the European Union Seventh Framework 
Program (FP7/2007-2013) under grant agreement No.312789 and
from the Foundation for Polish Science
through the Master/Mistrz program 3/2012. The research leading to these results has received funding from the  European Commission Seventh Framework Programme (FP/2007-2013)
under grant agreement No 283393 (RadioNet3)

%
%

\end{document}